\documentclass[11pt, oneside]{amsart}   	
\usepackage{geometry}                		
\usepackage{graphicx}				
\usepackage{amssymb}
\usepackage{amsmath}

\renewcommand\vec[1]{\mathbf{#1}}

\title{Evolutionary dynamics of cancer: from epigenetic regulation to cell population dynamics---mathematical model framework, applications, and open problems}
\author{Jinzhi Lei}
\begin{document}
\maketitle

\begin{center}
\begin{minipage}{15cm}
Zhou Pei-Yuan Center for Applied Mathematics, Tsinghua University, Beijing, China.
\end{minipage}
\end{center}
\vspace{0.5cm}

\textbf{Abstract}\ Predictive modeling of the evolutionary dynamics of cancer is a challenge issue in computational cancer biology. In this paper, we propose a general mathematical model framework for the evolutionary dynamics of cancer with plasticity and heterogeneity in cancer cells. Cancer is a group of diseases involving abnormal cell growth, during which abnormal regulations in stem cell regeneration are essential for the dynamics of cancer development. In general, the dynamics of stem cell regeneration can be simplified as a $\mathrm{G_0}$ phase cell cycle model, which lead to a delay differentiation equation. When cell heterogeneity and plasticity are considered, we establish a differential-integral equation based on the random transition of epigenetic states of stem cells during cell division. The proposed model highlights cell heterogeneity and plasticity, and connects the heterogeneity with cell-to-cell variance in cellular behaviors, \textit{e.g.}, proliferation, apoptosis, and differentiation/senescence, and can be extended to include gene mutation-induced tumor development. Hybrid computations models are developed based on the mathematical model framework, and are applied to the process of inflammation-induced tumorigenesis and tumor relapse after CAR-T therapy. Finally, we give rise to several mathematical problems related to the proposed differential-integral equation. Answers to these problems are crucial for the understanding of the evolutionary dynamics of cancer. 


\section{Introduction}

According to the World Health Organization (WHO), cancer has become the first ranking cause of death at ages below 70 years in 48 countries at 2015. Cancer has been a serious issue for public health. How cancers arise from normal tissues? Why is cancers so hard to treat? Despite many years researches, people know little about the evolutionary dynamics of cancer.  Multidisciplinary study is required to uncover the mechanism of cancer development.  Recent years, mathematical oncology--the use of mathematics in cancer research--has gained momentum with the rapid accumulation of data and applications of mathematical methodologies\cite{Rockne:2019cd}. 

Cancer is a group of diseases involving abnormal cell growth with the potential to invade or spread to other parts of the body. Hence, abnormal stem cell regeneration is essential in cancer development. Many mathematical models of the population dynamics have been widely studied in understanding how stem cell regeneration is modulated in different context \cite{Burns:1970tm,Dingli:2007ts,Hu:2012iy,Lander:2009fr,Mackey:1978vy,Mackey:2001ux,2008PLoSO...3.1591M,Mangel:2012ct,RodriguezBrenes:2011hw,Traulsen:2013dj,Zhou:2013ky}. In most models, the dynamics of a homogeneous cell pool or the lineage of several homogeneous subpopulations were formulated through a set of differential equations. Novel experimental techniques developed in recent years allow us to investigate the heterogeneity of cells at single cell level \cite{Chang:2008gua,Dykstra:2007dp,Gibson:2015ky,Hayashi:2008fu,Singer:2014eua,ZernickaGoetz:2009ita}. The heterogeneity is mostly originated  from random changes of epigenetic state at each cell cycle, and often induce cell plasticity and diversity which form the main sources of drug resistance in cancer therapy\cite{Graf:2002tj,LeMagnen:2018fx,Su:2017kp,Puram:2017ku}. Based on the stochastic transition of epigenetic state during cell division, mathematical models for the heterogeneous population dynamics can be formulated as discrete dynamical system or continuous differentiation-integral equations  \cite{2014PNAS..111E.880L,Situ:2017dg}. Moreover, hybrid models are often developed to simulate the process of tumor development and treatment of tumors as multicellular heterogeneous system of tumors\cite{Chamseddine:2019hz}.

In this paper, we present a general mathematical framework of modeling the evolutionary dynamics of cancer with plasticity and heterogeneity in the cells.  We first introduce a delay differential equation obtained from an age-structure model of $\mathrm{G_0}$ phase stem cell regeneration, and next propose  the mathematical formulations for heterogeneous population dynamics with  consideration of epigenetic state transition during cell divisions. Next, we introduce application of the proposed framework in case studied of cancer development. Finally, we give rise mathematical problems related to the proposed mathematical model. Answers to these problems are crucial for understanding the evolutionary dynamics of cancer. 

\section{A simple model--$G_0$ cell cycle model}
\label{sec:G0}

\begin{figure}[htbp]
\centering
\includegraphics[width=8cm]{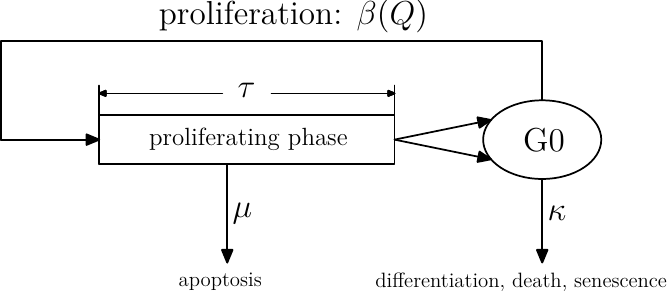}
\caption{The $\mathrm{G_0}$ model of stem cell regeneration. During stem cell regeneration, cells in the resting phase either enter the proliferating phase with a rate $\beta$, or be removed from the resting pool with a rate $\kappa$ due to differentiation, aging, or death. The proliferating cells undergo apoptosis with a probability $\mu$. }
\label{fig:ST:G0}
\end{figure}

First, we introduce a simple model of stem cell regeneration--the $G_0$ cell cycle model established by Burns and Tannock in 1970 \cite{Burns:1970tm}. This model assumed a resting phase ($\mathrm{G_0}$) between two cell cycles. Stem cells at cell cycling are classified into resting or proliferating phases (Figure \ref{fig:ST:G0}). During each cell cycle, a cell in the proliferating phase either undergoes apoptosis or divides into two daughter cells; however, a cell in the resting phase either irreversibly differentiates into a terminally differentiated cell or returns to the proliferating phase. Resting phase cells can also be removed from the stem cell pool due to cell death or senescence. 

Mathematically, the above process can be modeled by an age-structure model for cell numbers in the resting phase and proliferating phase. Let $s(t, a)$ for the number of stem cells at the proliferating phase, the age $a=0$ is the time of entry into the proliferative state, and $Q(t)$ the number of resting-phase stem cells. The above assumptions yield the following partial differential equations\cite{Burns:1970tm}
\begin{equation}
\label{eq:9}
\begin{array}{rcl}
\nabla s(t,a) &=& - \mu s(t, a), \quad (t > 0,  0 < a < \tau)\\
\dfrac{d Q}{d t} &=& 2 s(t, \tau) - (\beta(Q) + \kappa) Q,\quad  (t> 0).  
\end{array}
\end{equation}
Here  $\nabla = \partial/\partial t + \partial/\partial a$. The coefficient $2$ means that each mother cell generates two daughter cells after cell division.
The boundary condition at $a=0$ is as follows:
\begin{equation}
\label{eq:10}
s(t,0) = \beta(Q(t)) Q(t),
\end{equation}
and the initial conditions are 
\begin{equation}
\label{eq:11}
s(0, a) =g(a), \quad (0\leq a \leq \tau);\quad Q(0) = Q_0.
\end{equation}
Equations \eqref{eq:9}-\eqref{eq:11} give a general age-structure model of homogeneous stem cell regeneration. Integrating \eqref{eq:9}-\eqref{eq:10} through the method of characteristic line, and consider the long-term behavior ($t > \tau$), we obtain a delay differential equation model \cite{Jinzhi:2011bl}
\begin{equation}
\label{eq:5.15}
\dfrac{d Q}{d t} =  - (\beta(Q) + \kappa) Q + 2 e^{-\mu \tau}\beta(Q_\tau) Q_\tau,
\end{equation}
where $Q_\tau(t) = Q(t-\tau)$. This equation describes the general population dynamics of stem cell regeneration.  

The proliferation rate $\beta(Q)$ describes how cells regulate the self-renewal of stem cells through secreted cytokines, and is often given by a decrease function of form \cite{Bernard:2003ct,Mackey:1978vy}
\begin{equation}
\label{eq:6.23}
\beta(Q) = \beta_0 \dfrac{\theta^n}{\theta^n + Q^n}  + \beta_1.
\end{equation}
Here, a non-zero constant $\beta_1$ is included to represent the self-sustained growth signals of cancer cells\cite{Hanahan:2000hx}.

When $\beta(Q)$ is given by \eqref{eq:6.23}, the equation \eqref{eq:5.15} has a unique positive steady state if and only if 
\begin{equation}
\beta_0 + \beta_1 > \eta > \beta_1,\quad \mathrm{here}\ \eta = \dfrac{\kappa}{2 e^{-\mu \tau} - 1}.
\end{equation}
The steady state is given by 
\begin{equation}
Q^* = \eta \left[\dfrac{\beta_0 + \beta_1 - \eta}{\eta - \beta_1}\right]^{1/n}.
\end{equation}
Therefore, abnormal growth ($Q^* \to \infty$) may occur when $\eta \to \beta_1$, \textit{i.e.}, decreasing $\eta$ (by decreasing $\kappa$ or $\mu$) or increasing $\beta_1$. Hence, the simple $G_0$ cell cycle model \eqref{eq:5.15} implies three possible conditions for abnormal cell growth: dysregulation in the pathways of differentiation and/or senescence (decreasing $\kappa$), sustaining proliferative signaling (increasing $\beta_1$), and evasion of apoptosis (decreasing $\mu$). These conditions are well known hallmarks of cancer \cite{Hanahan:2000hx}.  

\section{A model with cell heterogeneity and plasticity}

\subsection{Modele equation}
\begin{figure}[htbp]
\centering
\includegraphics[width=8cm]{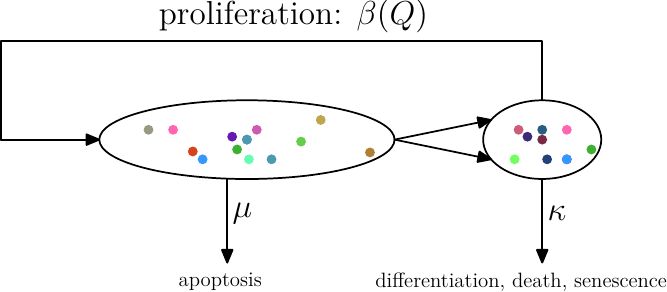}
\caption{Illustration of the model for heterogeneous stem cell regeneration. Similar to Figure \ref{fig:ST:G0}. The heterogeneity is represented by cells with different colors, and the hence the rate functions are dependent on the epigenetic state of the cell.}
\label{fig:model1}
\end{figure}

The above $G_0$ cell cycle model describes the population dynamics of homogeneous stem cell regeneration. However, cells are different by their epigenetic states, including the patterns of DNA methylations, nucleosome histone modifications, and the transcriptomics. These epigenetic states can change from mother cell to daughter cells during cell cycling through the re-distribution of modification markers, nucleosome assembling, or stochastic molecular partitions (Figure \ref{fig:model1}) \cite{Probst:2009iq,Schepeler:2014ir,SerraCardona:2018bz,Singer:2014eua,Takaoka:2014gg,Wu:2014gw}. 

To model the heterogeneity in stem cells, we introduce a valuable $\vec{x}$ (often a high dimensional vector) for the epigenetic state of cells, and $\Omega$ for the space of all possible epigenetic states in resting phase stem cells. The epigenetic state $\vec{x}$ represents intrinsic cellular states that may change during cell division. Let $Q(t,\vec{x})$ the number of stem cells at time $t$ in the resting phase with epigenetic state $\vec{x}$, the total cell number is given by 
\begin{equation}
\label{eq:Nt}
Q(t) = \int_{\Omega} Q(t, \vec{x})d\vec{x}.
\end{equation}

Proliferation of each cell is regulated by the signaling pathways that are dependent on extracellular cytokines released by all cells in the niche and the epigenetic state $\vec{x}$ of the cell\cite{Bernard:2003ct,Lander:2009fr,Mangel:2012ct}. Hence, the proliferation rate $\beta$ has a form $\beta(\hat{Q},\vec{x})$, where $\hat{Q}$ is the effective concentration of growth inhibition cytokines given by
\begin{equation}
\hat{Q}(t) = \int_{\Omega} Q(t,\vec{x}) \zeta(\vec{x}) d \vec{x},
\end{equation}
here $\zeta(\vec{x})$ is the rate of cytokines secretion. Moreover, the apoptosis rate $\mu$, the cell cycle duration $\tau$, and the differentiation rate $\kappa$ are dependent on the epigenetic state $\vec{x}$, and are denoted as $\mu(\vec{x})$, $\tau(\vec{x})$, and $\kappa(\vec{x})$, respectively. 

Moreover, to consider cell plasticity at each cell cycle, we introduce a transition function $p(\vec{x}, \vec{y})$ for the inheritance probability, which represent the probability that a daughter cell of state $\vec{x}$ comes from a mother cell of state $\vec{y}$. It is obvious that
$$\int_\Omega p(\vec{x}, \vec{y}) d \vec{x} = 1$$ 
for any $\vec{y}\in \Omega$. 

Now, similar to \eqref{eq:9}, when stem cell heterogeneity is include, we obtain the corresponding age-structure model equation
\begin{equation}
\label{eq:m14}
\begin{array}{rcl}
\displaystyle \nabla s(t,a,\vec{x}) &=& \displaystyle - \mu(\vec{x}) s(t, a,\vec{x}), \quad (t > 0,\quad  0 < a < \tau(\vec{x}))\\
\displaystyle \dfrac{\partial Q(t,\vec{x})}{\partial t} &=& \displaystyle 2 \int_\Omega s(t, \tau(\vec{y}), \vec{y})p(\vec{x}, \vec{y})d \vec{y} - (\beta(\hat{Q}(t),\vec{x}) + \kappa(\vec{x})) Q(t,\vec{x}),\quad  (t> 0).  
\end{array}
\end{equation}
and
$$s(t,0,\vec{x}) = \beta(\hat{Q}(t), \vec{x})Q(t,\vec{x}),\quad \hat{Q}(t) = \int_\Omega Q(t,\vec{x}) \zeta(\vec{x}) d \vec{x}.$$
Here we note that $\vec{x}$ can be considered as a parameter for the first equation, hence, we can apply the characteristic line method and have
$$s(t,\tau(\vec{x}), \vec{x}) = \beta(\hat{Q}(t-\tau(\vec{x})), \vec{x}) Q(t-\tau(\vec{x}), \vec{x})e^{-\mu(\vec{x}) \tau(\vec{x})}.$$
Thus, substituting $s(t, \tau(\vec{x}), \vec{x})$ into the second equation in \eqref{eq:m14}, we obtain the following delay differential-integral equation
\begin{equation}
\label{eq:6.26}
\left\{
\begin{array}{rcl}
\displaystyle\dfrac{\partial Q(t,\vec{x})}{\partial t} &=& \displaystyle-Q(t,\vec{x}) (\beta(\hat{Q},\vec{x})  + \kappa(\vec{x}))\\
&&\displaystyle{} + 2 \int_{\Omega} \beta(\hat{Q}_{\tau(\vec{y})},\vec{y}) Q(t-\tau(\vec{y}),\vec{y}) e^{-\mu(\vec{y})\tau(\vec{y})} p(\vec{x},\vec{y}) d\vec{y},\\
\hat{Q}(t)&=&\displaystyle\int_{\Omega} Q(t,\vec{x}) \zeta(\vec{x}) d\vec{x},
\end{array}
\right.
\end{equation}
Here $\hat{Q}_\tau = \hat{Q}(t-\tau)$.

\begin{figure}[htbp]
\centering
\includegraphics[width=12cm]{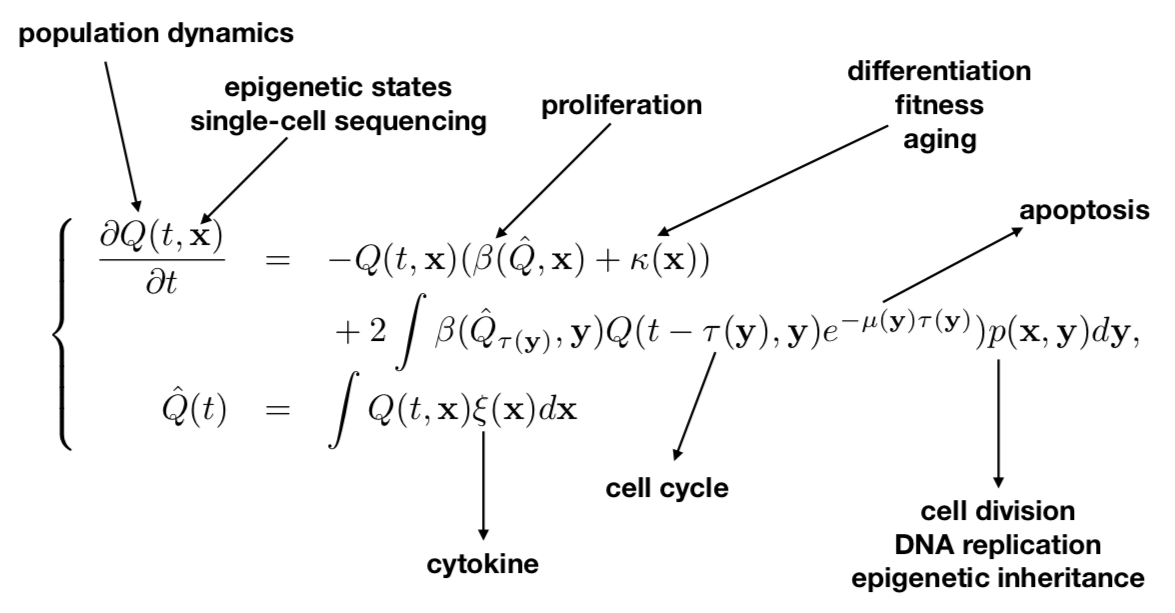}
\caption{Framework of the the mathematical model of heterogeneous stem cell regeneration.}
\label{fig:4}
\end{figure}

The equation \eqref{eq:6.26} gives the basic dynamical equation of heterogenous stem cell regeneration with epigenetic transition. Biologically, the equation \eqref{eq:6.26} connects different scale components(Figure. \ref{fig:4}): the gene expressions at single cell level ($\vec{x}$), the population dynamic properties ($\beta(\hat{Q},\vec{x}), \kappa(\vec{x})$, and $\mu(\vec{x})$), cell cycle ($\tau(\vec{x})$), and epigenetic modification inheritance ($p(\vec{x}, \vec{y})$). Thus, this equation provides a framework of mathematical model for stem cell regeneration, and can be applied to different problems related to cell regeneration, such as development, aging, tumor development, \textit{etc}. For detail discussions of \eqref{eq:6.26}, referred to \cite{Lei:2019vd}. 

\subsection{The transition function $p(\vec{x}, \vec{y})$}

In the above equation, the transition function $p(\vec{x}, \vec{y})$ is important to connect cell heterogeneity with plasticity. However, biologically we cannot measure this function directly, and do not know the possible form neither.  Here, we propose a possible form of the transition function through numerical simulation based on a computational model of the inheritance of histone modification\cite{Huang:2017jr,Huang:2019hn}.   

Let the epigenetic state $x$ ($0 \leq x \leq 1$) represent the fraction of nucleosomes (in a segment of DNA sequences), and $p(x,y)$ the probability of a daughter cell with state $x$ given the mother cell of state $y$. Simulations based on a computation model of histone modification/transition suggest that the probability $p(x,y)$ can be approximately described by a conditional Beta-distribution
$$p(x, y) = \dfrac{x^{a(y) - 1} (1-x)^{b(y) - 1}}{B(a(y), b(y))},\quad B(a, b) = \dfrac{\Gamma(a) \Gamma(b)}{\Gamma(a+b)}.$$
Here, the Beta-distribution depends on two shape parameters $a$ and $b$, which are in turn dependent on the state $y$ of the mother cell. Hence, the Beta-distribution can be quite general with different definitions of the functions $a(y)$ and $b(y)$.

In general, when the epigenetic state $\vec{x} = (x_1, \cdots, x_n)$ that includes multiple variables, we can extend the above Beta-distribution by the multiply rule
$$
p(\vec{x}, \vec{y}) = \sum_{i=1}^n p_i(x_i, \vec{y}),
$$
where
$$p_i(x_i, \vec{y}) = \dfrac{x_i^{a_i(\vec{y}) - 1} (1-x_i)^{b_i(\vec{y}) - 1}}{B(a_i(\vec{y}), b_i(\vec{y}))}.
$$
To determine the functions $a_i(\vec{y})$ and $b_i(\vec{y})$, if we write the mean and variance of $x_i$, given the state $\vec{y}$, as
$$\mathrm{E}(x_i | \vec{y}) = \phi_i(\vec{y}), \quad \mathrm{Var}(x_i | \vec{y}) = \dfrac{1}{1 + \eta_i(\vec{y})} \phi_i(\vec{y}) (1 - \phi_i(\vec{y})),$$
through predefined function $\phi_i(\vec{y})$ and $\eta(\vec{y})$, the shape parameters are given by
$$
a_i(\vec{y}) = \eta_i(\vec{y})\phi_i(\vec{y}), \quad b_i(\vec{y}) = \eta_i(\vec{y}) (1 - \phi_i(\vec{y})).
$$
Here, the functions $\phi_i(\vec{y})$ and $\eta_i(\vec{y})$ always satisfy
$$0 < \phi_i(\vec{y}) < 1, \quad \eta_i(\vec{y}) > 0.$$

As an example, if we consider a situation of one epigenetic state $x\ (0 \leq x\leq 1)$ which is analogous to the stemness of a cell \cite{Malta:2018ic}. The stemness may affect cell proliferation and differentiations so that the rates $\beta$ and $\kappa$ are dependent on the epigenetic state $x$. Moreover, we assume that $\xi(x)\equiv 1$ so that
\begin{equation}
\hat{Q}(t)  = \int_0^1 Q(t,x) d x.
\end{equation}
In this case, we have a one dimensional differential-integral equation
\begin{equation}
\label{eq:hsc1}
\dfrac{\partial Q(t, x)}{\partial t} = - Q(t, x) (\beta(\hat{Q}(t), x) + \kappa(x)) + 2 e^{-\mu \tau}\int_0^1\beta(\hat{Q}(t-\tau), y) Q(t-\tau, y)  p(x,y) d y.
\end{equation} 
The inheritance probability function $p(x,y)$ is defined as
\begin{equation}
p(x,y) = \dfrac{x^{a(y)}(1-x)^{b(y)}}{B(a(y), b(y))},\quad a(y) = \eta(y) \phi(y), b(y) = \eta(y) (1 - \phi(y))
\end{equation}
with pre-defined functions $\phi(y)$ and $\eta(y)$. 

Specifically, we can define \cite{Lei:2019vd}
\begin{equation}
\label{eq:b14}
\beta(\hat{Q},x) = \bar{\beta} \times \dfrac{a_1 x + (a_2 x)^6}{1 + (a_3 x)^6} \dfrac{\theta}{\theta + \hat{Q}} + \beta_1,
\end{equation}
and 
\begin{equation}
\label{eq:k15}
\kappa(x) = \kappa_0 \times \dfrac{1}{1 + (b_1 x)^6}.
\end{equation}
The functions $\beta$ and $\kappa$ defined by \eqref{eq:b14} and \eqref{eq:k15} are taken so that a cell has low proliferation rate if the stemness if either high or low, and large proliferation rate if the stemness is intermediate, and the differentiation rate decreases with the stemness. 

\subsection{Model cancer development with genetic heterogeneity}
In the above equations, we only consider heterogeneous in epigenetic states. However, gene mutations are important for cancer development, which form intratumoral heterogeneity of cancer cells. The cells may have diverse combinations of mutations. To model the genetic heterogeneity, we extend \eqref{eq:6.26} to include mutant types. Assuming that there are $m$ genetic types, we let $Q_i(t, \vec{x})$ the cell population with genetic type $i$ ($1\leq i \leq m$) and epigenetic state $\vec{x}$, and $p_{i,j}$ the mutation rate from genetic type $i$ to type $j$. Moreover, we assume that mutations only happen during cell division (DNA replication). Thus, when gene mutations are include, the equation \eqref{eq:6.26} is rewritten as
\begin{eqnarray}
\label{eq:m16}
\dfrac{\partial Q_i(t, \vec{x})}{\partial t} &=& - Q_i(t, \vec{x}) (\beta_i(\hat{Q}, \vec{x}) + \kappa_i(\vec{x}))\\
&&{} + 2 (1 - \sum_{j\not=i} p_{i,j}) \int_\Omega \beta_i(\hat{Q}_{\tau_i(\vec{y})}, \vec{y}) Q_i(t - \tau(\vec{y}), \vec{y}) e^{-\mu_i(\vec{y}) \tau_i(\vec{y})} p(\vec{x}, \vec{y}) d \vec{y}\nonumber\\
&&{} + 2 \sum_{j\not=i} p_{j,i} \int_\Omega \beta_j(\hat{Q}_{\tau(\vec{y})}, \vec{y}) Q_j(t - \tau(\vec{y}),\vec{y}) e^{-\mu_j(\vec{y})\tau_j(\vec{y})} p(\vec{x}, \vec{y}) d \vec{y},\nonumber\\
&&{} \qquad (1 \leq i \leq m)\nonumber
\end{eqnarray} 
where
$$\hat{Q}(t) = \sum_{i=1}^m \int_\Omega Q_i(t,\vec{x}) \xi(\vec{x}) d \vec{x}.$$

Equation \eqref{eq:m16} gives a general equation to include gene mutations. Given the mutant type combination network $p_{i,j}$, this equation describes the evolutionary dynamics induced by gene mutations. Mathematically, the mutation may affect the cell behavior parameters, so that the rate constants $\beta, \kappa, \mu$, and $\tau$ are dependent on the mutant types.  

\section{Applications to the dynamics of cancer development}

\subsection{Hybrid computational model of multicellular tissues}
The equations \eqref{eq:6.26} and \eqref{eq:m16} provide general mathematical frameworks to model stem cell regeneration when heterogeneity and plasticity in epigenetic or genetic states are included. These frameworks can be used to described many biological processes associated with stem cell regeneration, including development, aging, and cancer evolution. Nevertheless, many mathematical problems associated with the model remain unsolved (to be detailed latter). Both equations \eqref{eq:6.26} and \eqref{eq:m16} include integrals over all epigenetic states, and it is expansive to solve the equations numerically when high dimensional epigenetic states are considered. Thus, in applications, we often develop hybrid computational models for multicellular tissues based on the above frameworks. Hybrid models are often used to simulate tumor development as complex multicellular heterogeneous systems\cite{Chamseddine:2019hz}. Classical hybrid models combine discrete equations to describe individual cells and continuous equations for microenvironment factors or intercellular components. Now, the field of tumor modeling have reached into other mathematical areas and combined continuous or discrete models with concepts from fluid dynamics, game theory, machine learning, or optimization methods to provide more powerful predictive models\cite{Chamseddine:2019hz}. 

The above mathematical framework suggest  a hybrid model that combines discrete stochastic process for the epigenetic/genetic state of individual cells with continues model of cell population growth (here, we state the numerical scheme for epigenetic state heterogeneity, and it is similar when genetic heterogeneity is included). In this model, a multicellular system is represented by a collection of epigenetic states in each cell as $\Omega_t = \left\{[C_i(\vec{x}_i)]_{i=1}^{Q(t)}\right\}$ (here $Q(t)$ represents the number of resting-phase stem cells at time $t$). During a time interval $(t, t + dt)$, each cell ($C_i(\vec{x}_i)$) undergoes proliferation, apoptosis, or terminal differential with a probability given by the kinetic rates (with probability of proliferation, apoptosis, or differentiation given by $\beta(\hat{Q}, \vec{x}_i)) dt, \mu(\vec{x}_i) dt$, or $\kappa(\vec{x}_i) dt$). Thus, the probabilities of different cell behaviors are dependent on the epigenetic state of each cells. The total cell number $Q(t)$ changes after a time step $dt$ in accordance with the behaviors of all cells. When a cell undergo proliferation, the epigenetic state of daughter cells change randomly according to the transition function $p(\vec{x}, \vec{y})$. In this hybrid model, all detail molecular interactions are hidden into the kinetics rates and the transition function. Moreover, we can also include stochastic process of differential equations for the signaling dynamics within one cell cycle, as well as microenvironmental variables that may depend on the cell behavior of all cells. The propose hybrid model can be implement by single-cell-based models through GPU architecture\cite{Song:2018kl}. 

\subsection{Application to inflammation-induced tumorigenesis}
Chronic inflammation is a serous risk factor for many cancers. Infection-driven chronic inflammation is linked to approximately 15\% of the global cancer burden\cite{Elinav:2013hv,Martel2012,Parkin2006}. Cancer risk increases strongly with the duration and extent of chronic inflammation\cite{Grivennikov:2010dk,Clevers2004}.  However, the routes from inflammation to cancer are poorly understood.  

In \cite{Guo:2017ix}, following the above mathematical model framework, we developed a computation model for inflammation-induced tumorigenesis that combines the major processes responsible for stem cell regeneration, inflammation, and metabolism-immune balance. In this model, the population dynamics arise from the dynamics of individual cells, each of which is based on a model of the cell-cycle dynamics, and the proliferation rate is dependent on the population size.  Each cell is associated with individual genetic heterogeneity due to DNA damage and pathway mutations. DNA damage may occur to a cell undergoing cell division, and trigger the processes of DNA damage repair and cell-cycle arrest or DNA damage-induced apoptosis when the damaged loci are not successively repaired. Non-repaired cells can escape from damage-induced apoptosis and reenter the cell cycle such that over time, damage loci accumulate and eventually induce functional gene mutations in the specific pathways. Mutant cells in the resting phase cell pool can be cleared by the immune system due to the metabolism-immune balance.  The inflammatory microenvironment specifically affects the processes of proliferation, apoptosis, and the DNA damage response. Gene mutations were not considered explicitly in the model; however, the mutations were identified by their effects on the relevant physiologic processes. Mutations to eight pathways were considered (Table \ref{tab:1}), each type mutation occur with a given probability at each cell cycle. When a mutation occur to a cell, the corresponding parameter value changes (either up-regulated or down-regulated) in the cell. Each cell may have different mutant types. 

\begin{table}[htbp]
\caption{Pathway mutations considered in the model of inflammation-induce tumorigenesis\cite{Guo:2017ix}.}
\begin{center}
\begin{tabular}{|c|c|}
\hline
\textbf{Symbol} & \textbf{Description}\\
\hline
$\Delta$Prolif & Cell proliferation rate\\
\hline
$\Delta$FSProlif & Feedback strength to cell proliferation\\
\hline
$\Delta$Diff & Cell differentiation rate\\
\hline
$\Delta$Apop & Cell apoptosis rate\\
\hline
$\Delta$Damage & Probability of DNA damage induction\\
\hline
$\Delta$Repair & DNA damage repair efficiency\\
\hline
$\Delta$Escape & Probability of DNA damage-induced cell escape\\
\hline
$\Delta$MIB & Probability of metabolism-immune balance\\
\hline
\end{tabular}
\end{center}
\label{tab:1}
\end{table}%

Based on model simulations, we are able to reproduce the process of inflammation-induced tumorigenesis: from normal to precancerous, and from precancerous to malignant tumors.  According to model simulations, starting from serve inflammation, most patients develop to precancerous with insignificant increase in the cell population in 10 years, and a few patients may further develop to malignant tumors with signifiant increase in cell number in latter stages. Moreover, further analysis shown that mutations to the four pathways, proliferation, apoptosis, differentiation, and metabolism-immune balance, are crucial for cancer development, and there are multiple pathways of tumorigenesis\cite{Guo:2017ix}. 

\subsection{Application to tumor relapse after CAR-T therapy}

CAR-T therapy targeting CD19 has been proved to be an effective therapy for B cell acute lymphoblastic leukemia (B-ALL). The majority of patients achieve a complete response following a single infusion of CD19-targeted CAR-T cells; however, many patients suffer relapse after therapy, and the underlying mechanism remains unclear. 

In a recent study\cite{Zhang:2019fs}, we applied second-generation CAR-T cells to mice injected with leukemic cells; 60\% of the mice relapse within 3 months, and the relapsed tumors retained CD19 expression but exhibited a profound increase in CD34 transcription. Based on these observations, we develop a single-cell based computation model for the heterogeneous response of the tumor cells to the CAR-T treatment. 

In the model, we introduced key assumptions that CAR-T induced tumor cells to transition to hematopoietic stem-like cells (by promoted CD34 expression) and myeloid-like cells (my promoted CD123 expression) and hence escape of CAR-T targeting. In the model, each cell was represented by the epigenetic state of marker genes CD19, CD22, CD34, and CD123, which play important roles in the CD19 CAR-T cell response and cell lineage dynamics. The proliferation rate $\beta$ and differentiation rate $\kappa$ depend on CD34 expression level through
\begin{eqnarray*}
\beta &=&\beta_0 \dfrac{\theta}{\theta + N} \times \dfrac{5.8 [\mathrm{CD34}] + (2.2 [\mathrm{CD34}])^6}{1 + (3.75 [\mathrm{CD34}])^6},\\
\kappa &=& \kappa_0 \dfrac{1}{1 + (4.0 [\mathrm{CD34}])^6}.
\end{eqnarray*} 
Here $N$ means the total cell number. The apoptosis rate $\mu$ includes a basal rate $\mu_0$ and a rate associated with the CAR-T signal
\begin{eqnarray*}
\mu &=& \mu_0 + \mu_1 \times \mathrm{Signal},\\
\mathrm{Signal} &=& f([\mathrm{CD34}], [\mathrm{CD123}]) \dfrac{\gamma_{19}[\mathrm{CD19}]}{1 + \gamma_{19}[\mathrm{CD19}]+\gamma_{22}[\mathrm{CD22}]}R(t),\\
f([\mathrm{CD34}], [\mathrm{CD123}]) &=& \dfrac{1}{(1 + ([\mathrm{CD34}]/X_0)^{n_0}) (1 + ([\mathrm{CD123}]/X_1)^{n_1})}.
\end{eqnarray*}
Here $R(t)$ is the predefined CAR-T activity. The expression levels of marker genes changed randomly following the transition probability of Beta-distributions, and the shape parameters were dependent on state of mother cells and the CAR-T signal. For example, given  the expression level of CD34 at cycle $k$ as $u_k$, the expression level at cycle $k+1$ (denoted by $u_{k+1}$) is a Beta-distribution random number with probability density function 
$$p(u, u_k) = \dfrac{u^a (1-u)^b}{B(a,b)},\quad B(a,b) = \dfrac{\Gamma(a)\Gamma(b)}{\Gamma(a+b)},$$
with the shape parameters $a$ and $b$ dependent on the average and variance of $u_{k+1}$. When 
$$\mathrm{E}(u_{k+1}) = \phi(u_k), \quad \mathrm{Var}(u_{k+1}) = \dfrac{1}{1+m}\phi(u_k) (1 - \phi(u_k)),$$
then
$$a = m \varphi(u_k),\quad b = m (1 - \varphi(u_k)). $$
In the model, we assume the average as
$$\phi(u_k) = 0.08 + 1.06 \dfrac{(\alpha_{34} u_k)^{2.2}}{1 + (\alpha_{34} u_k)^{2.2}},$$
and let
$$\alpha_{34}  = 1.45 + 0.16\times [\mathrm{CD19}] + \alpha_{34} \times \mathrm{Signal}$$
to represent the promotion of CD34 expression by CD19 and the CAR-T signal. For details of the model, referred to \cite{Zhang:2019fs}.  

Model simulations nicely reproduced experimental results, and predicted that CAR-T cell induced cell plasticity can lead to tumor tumor relapse in B-ALL after CD19 CAR-T treatment. Simulations and mouse experiments further indicated that CD19 positive relapse could be prevented by the combined administration of CD19- and CD123- targeting CAR-t cells administered at specific ratios.   

\section{Mathematical problems}
The proposed mathematical framework \eqref{eq:6.26} is a delay differential-integral equation that contains non-local transitions between different epigenetic states. This type equation was not seen in most physical problems because of the principle of locality. Mathematically, there are many basic problems remain open for the equation \eqref{eq:6.26}. Here, I discuss a few of them that are basic and important for our understanding of the biological process of cancer development. 

For the simplicity, we omit the delay, and consider the equation with only one dimension epigenetic state. Hence, we have the following equation
\begin{equation}
\label{eq:m17}
\begin{array}{rcl}
\displaystyle\dfrac{\partial Q(t, x)}{\partial t} &=& \displaystyle-Q(t, x)(\beta(\hat{Q}, x) + \kappa(x)) + 2 \int_{\Omega} \beta(\hat{Q}, y) Q(t, y) e^{-\mu(y)} p(x,y) dy\\
\displaystyle\hat{Q}(t) &=& \displaystyle \int_\Omega Q(t,x) \zeta(x) d x.
\end{array}
\end{equation}
Here $x\in \Omega \subset \mathbb{R}^+$, and always assume
\begin{equation}
\label{eq:m18}
\beta(x) \geq 0, \kappa(x) \geq 0, \mu(x) \geq 0, \zeta(x) \geq 0,  0 \leq p(x, y) \leq 1,
\end{equation}
and 
\begin{equation}
\label{eq:m19}
\int_\Omega p(x,y) d x = 1,\quad \forall y \in \Omega.
\end{equation}

Here, we note that by omitting the delay, we do not simply set $\mu = 0$ in the equation \eqref{eq:6.26}, but only omit the delays in $Q(t-\tau(\vec{y}), \vec{y})$ and $\hat{Q}_{\tau(\vec{y})}$, and replace $e^{-\mu(\vec{y})\tau(\vec{y})}$ by $e^{-\mu(\vec{y})}$ with $\mu(\vec{y})$ the apoptosis rate during the proliferating phase. 

\subsection{Question 1: Existence and uniqueness of the steady state solution}
To consider the steady state solution of \eqref{eq:m17}, let $Q(t, x) = Q(x)$ the steady state, then 
\begin{equation}
\label{eq:m20}
\left\{
\begin{array}{rcl}
\displaystyle 0 &=&\displaystyle -Q(x) (\beta(\hat{Q}, x) + \kappa(x)) + 2 \int_\Omega \beta(\hat{Q}, y) Q(y) e^{-\mu(y)} p(x, y) d y\\
\displaystyle \hat{Q} &=&\displaystyle \int_\Omega Q(x) \zeta(x) dx
\end{array}.
\right.
\end{equation}
Substituting $\hat{Q}$ into the first equation, $Q(x)$ satisfies the eigenvalue problem
\begin{equation}
L_{\hat{Q}}[Q(x)] = 2 \int_\Omega \beta(\hat{Q}, y) e^{-\mu(y)} p(x,y) Q(y) d y - (\beta(\hat{Q}, x) + \kappa(x)) Q(x) = 0.
\end{equation}
Thus, the problem of existence and uniqueness of the steady state solution is reduced to a problem of finding a positive eigenvalue $\hat{Q}$ of the operator $L_{\hat{Q}}$ so that the corresponding eigenfunction $Q(x)$ is non-negative for all $x\in \Omega$.  If such eigenvalue $\hat{Q}$ exists, the solution of \eqref{eq:m20} is given by rescaling the corresponding eigenfunction according to the second equation in \eqref{eq:m20}.

In the case of finite discrete epigenetic state, and when the proliferation rate $\beta$ is independent to the epigenetic state, the steady state issue was discussed in \cite{Situ:2017dg}. In particular, if $\beta$ is independent to the epigenetic state, the eigenvalue problem can be rewritten as
\begin{equation}
A[Q] = \dfrac{1}{\beta(\hat{Q})} Q,
\end{equation}
where $A$ is a linear operator defined as
\begin{equation}
A[Q] = \dfrac{1}{\kappa(x)}\left[2\int_{\Omega} e^{-\mu(y)}p(x,y)Q(y)dy - Q(x)\right].
\end{equation}
Thus, $\dfrac{1}{\beta(\hat{Q})}$ is a positive eigenvalue of the operator $A$. In the case of finite discrete epigenetic state, the existence for such eigenvalue can be obtained from the Perron-Frobenius theorem.

If the transition probability $p(x,y)$ is independent to the state of the mother cell, so that $p(x,y) = p(x)$, the uniqueness and stability of the steady state were further discussed in \cite{Situ:2017dg}. However, it remains open questions for general cases.

Biologically, the states of normal and precancerous correspond to stable steady states under different conditions. Hence, mathematically identify the existence and stability of steady states are important for our understanding of the persistence of different states.

\subsection{Question 2: Entropy problem}

Let  
\begin{equation}
Q(t) = \int_\Omega Q(t,x)
\end{equation}
for the total cell number at time $t$, and 
\begin{equation}
f(t,x) = \dfrac{Q(t,x)}{Q(t)}
\end{equation} 
for the fraction of cells with given epigenetic state $x$. The entropy of the multicellular system at time $t$ is defined as
\begin{equation}
E(t) = -\int_{\Omega} f(t,x) \log f(t, x) d x.
\end{equation}

A tedious calculation gives
\begin{eqnarray}
\label{eq:m27}
\dfrac{\partial f(t,x)}{\partial t} &=& - f(t,x) \int_{\Omega} f(t,y) \left((\beta(\hat{f}Q,x) + \kappa(x)) - (\beta(\hat{f}Q,y)  + \kappa(y))\right)dy\\
&&{} + 2 \int_{\Omega} f(t,y) \beta(\hat{f}Q,y) e^{-\mu(y)}\left(p(x,y) - f(t,x)\right) d y,\nonumber\\
\label{eq:m28}
\dfrac{d Q}{d t} &=& Q \int_{\Omega} f(t, x) \left(\beta(\hat{f}Q,x) (2 e^{-\mu(x)} - 1) - \kappa(x)\right) d x,\\
\label{eq:m29}
\hat{f} &=& \int_{\Omega} f(t,x) \zeta(x) d x.
\end{eqnarray}
The derivative of the entropy $E(t)$ is given by 
\begin{equation}
\dfrac{d E}{d t} = - \int_{\Omega} \dfrac{\partial f(t,x)}{\partial t} \left(1 + \log f(t, x)\right) d x. 
\end{equation}
Here, we replace $\hat{Q}$ with $\hat{f} Q$, and have $\hat{f} = 1$ when $\zeta(x)\equiv 1$.

The entropy problem asks: how can we decompose the rate of entropy change into entropy production rate ($epr$) and entropy dissipate rate ($edr$) so that
\begin{equation}
\dfrac{d E}{d t} = epr - edr,
\end{equation}
and both $epr$ and $edr$ are alway non-negative.  

Solutions to this problem is essential to answer the question of how the entropy change along the process of tissue development. In particular, is cancer development a process of entropy increasing\cite{Hanselmann:2016kg,Tarabichi:2013ev}? 

In the case of homogeneous stem cells, all cells belong to the same epigenetic state, and hence $f(t,x) = 1$, and the entropy $E(t) \equiv 0$. Thus, the entropy do not change over the development process. 

If the epigenetic state transition is omitted, so that $p(x, y) = \delta(x-y)$, defining
\begin{equation}
\gamma(q, x) = \beta(q, x)(2 e^{-\mu(x)} - 1) - \kappa(x)
\end{equation}
as the net production rate of cells with epigenetic state $x$, we have
\begin{equation}
\dfrac{\partial f(t, x)}{\partial t} = f(t, x) \gamma(\hat{f} Q, x) - f(t, x)\int_\Omega f(t, y) \gamma(\hat{f} Q, y) d y,
\end{equation} 
and 
\begin{equation}
\label{eq:m33}
\dfrac{d E}{d t} = \Delta E_1 - \Delta E_2, 
\end{equation}
where
\begin{equation}
\Delta E_1 = - \int_{\Omega} \gamma(\hat{f} Q, x) f(t,x) \log f(t, x) d x,\quad  \Delta E_2 =  E(t) \int_{\Omega} \gamma(\hat{f} Q, x) f(t,x) d x.
\end{equation}

From \eqref{eq:m33}, for the particular situation that $\gamma(\hat{f} Q, x) > 0$ for all $x\in \Omega$, we have $\Delta E_1 > 0$ and $\Delta E_2 > 0$.  In this case, $\Delta E_1$ gives the entropy production rate, and $\Delta E_2$ gives the entropy dissipation rate.

For the general situation, it is not know how the entropy production rate and dissipation rate should be defined, and under which situation the development process is entropy increasing.

\subsection{Question 3: Inverse problem}
In the propose model, the transition probability function $p(x,y)$ is usually not known, and cannot be measured directly from experiments. Hence, it is a challenge issue to estimate this transition function indirectly from experimental data.  

In the equations \eqref{eq:m27}-\eqref{eq:m28}, for the situation of cancer development, we can estimate the cell number $Q(t)$ (or tumor volume) and the fraction of cells $f(t, x)$ through various methods, such as imaging, liquid biopsies, biomarker measurement, single-cell sequencing, \textit{etc}.  Hence, the inverse problem of obtaining the rate functions $\beta, \kappa, \mu$, and $\zeta$, and the transition function $p(x, y)$ is essential for the development of personalized predictive model  for cancer development. It was proposed that the merger of mechanistic and machine learning models has became the future of personalization in mathematical oncology\cite{Rockne:2019cd}.  In this way, we are trying to merger the propose model with the unwieldy multitude of dispersed data (imaging, tissue, blood, molecular) to estimate the personalized parameter and to generate optimal clinical decisions for each patient.   

\subsection{Question 4: Local state transition}

In the equations \eqref{eq:m17} or \eqref{eq:m17}-\eqref{eq:m28}, the epigenetic transition function $p(x, y)$ is global. Here, we assume that the epigenetic state can only have local transition, \text{i.e.}, 
\begin{equation}
\label{eq:m35}
p(x, y) = \varphi(y-x)
\end{equation}
so that $\varphi(r) > 0$ only when $|r| < \epsilon \ll 1$.  The function $\varphi(r)$ satisfies
\begin{equation}
\int_{-\infty}^{+\infty} \varphi(r) d r = 1,\quad \varphi(r) \geq 0, \quad \forall r \in \mathbb{R}.
\end{equation}
We further let
\begin{equation}
a = \int_{-\infty}^{+\infty} r \varphi(r) d r, \quad D = \int_{-\infty}^{+\infty} r^2 \varphi(r) d r. 
\end{equation}
Substituting \eqref{eq:m35} into \eqref{eq:m17}, and expand the function $\varphi$ to the second order approximation, we obtain the following close form differential-integral equation
\begin{eqnarray}
\label{eq:m38}
\dfrac{\partial f(t, x)}{\partial t} &=& \dfrac{\partial\ }{\partial x} \left[\left(D\dfrac{\partial\ }{\partial x} + 2 a\right)\left(\beta(\hat{f} Q, x) e^{-\mu(x)}f(t, x)\right)\right]\\
&&{} + f(t, x) \left(\gamma(\hat{f} Q, x) - \int f(t, y) \gamma(\hat{f} Q, y) d y\right),\nonumber\\ 
\label{eq:m39}
\dfrac{d Q}{d t} &=& Q \int_\Omega f(t, x) \gamma(\hat{f} Q, x) d x\\
\label{eq:m40}
\hat{f} &=& \int_\Omega f(t, x) \zeta(x) d x.
\end{eqnarray}

From \eqref{eq:m38}-\eqref{eq:m40}, the above questions of steady state solution, entropy problem, and inverse problem can also be formulated as the problem with local transition. For example, at the steady state, we have
$$\int_\Omega f(t, x) \gamma(\hat{f} Q, x)  dx = 0, $$
and the total number $Q(t) \equiv Q^*$. Hence, the steady state solution ($f(t, x) = f(x)$) satisfies a second order differential equation
\begin{equation}
\dfrac{d\ }{d x} \left[\left(D\dfrac{d\ }{d x} + 2 a\right)\left(\beta(\lambda, x) e^{-\mu(x)}f(x)\right)\right] + \gamma(\lambda, x) f(x)  = 0,
\end{equation}
with boundary condition (here $\Omega \subseteq \mathbb{R}$)
$$f(x) \geq 0,\quad \int_{\Omega} f(x) d x = 1,\quad Q^* \int_\Omega f(x) \zeta(x) d x = \lambda.$$
This equation defines a  nonlinear eigenvalue problem for the eigenvalue $\lambda = \hat{f} Q^*$. 

\section{Discussions}

Cancer is a group of diseases with abnormal cell growth. This paper provides a general mathematical framework to describe the dynamics of heterogeneous stem cell regeneration, and introduce applications of the model to  the study of cancer development. The model highlights the cell heterogeneity and plasticity, and provides a connections between heterogeneity with cellular behavior, \textit{e.g.}, proliferation, apoptosis, and differentiation/senescence. We also extend the model to include gene mutations that are important for cancer development. Nevertheless, many other factors may also play important roles in cancer development, and are not included in the proposed model, such as the cell-to-cell interactions, immune cells, microenvironment, spatial information, \textit{etc.} The current equation should be extended to included these factors for a more complete model.

The proposed differential-integral equation model provides a general mathematical formulation for the process of stem cell regeneration. This equation integrates different scale interactions from epigenetic regulation to cell population dynamics. Mathematically, many basic properties of the proposed differential-integral equation are not known and remain challenge issues in future studies. Here we propose several mathematical problems that are basic and important for our understanding of cancer development. We hope that these equations can be the guideline for related studies in the future. 

Finally, the propose model is only a mathematical framework for stem cell regeneration with heterogeneity and plasticity, detail implementation of the model should be subjected to specific biological problems. 

\section*{Acknowledgements}
This research is supported by the National Natural Science Foundation of China (NSFC91730101 and 11831015).


\begin{thebibliography}{10}

\bibitem{Bernard:2003ct}
Samuel Bernard, Jacques B{\'e}lair, and Michael~C Mackey.
\newblock {Oscillations in cyclical neutropenia: new evidence based on
  mathematical modeling}.
\newblock {\em J Theor Biol}, 223(3):283--298, August 2003.

\bibitem{Burns:1970tm}
F~J Burns and I~F Tannock.
\newblock {On the existence of a G0-phase in the cell cycle}.
\newblock {\em Cell Proliferation}, 3(4):321--334, 1970.

\bibitem{Chamseddine:2019hz}
Ibrahim~M Chamseddine and Katarzyna~A Rejniak.
\newblock {Hybrid modeling frameworks of tumor development and treatment.}
\newblock {\em Wiley Interdiscip Rev Syst Biol Med}, page e1461, July 2019.

\bibitem{Chang:2008gua}
Hannah~H Chang, Martin Hemberg, Mauricio Barahona, Donald~E Ingber, and Sui
  Huang.
\newblock {Transcriptome-wide noise controls lineage choice in mammalian
  progenitor cells.}
\newblock {\em Nature}, 453(7194):544--547, May 2008.

\bibitem{Clevers2004}
Hans Clevers.
\newblock {At the crossroads of inflammation and cancer}.
\newblock {\em Cell}, 118(6):671--674, September 2004.

\bibitem{Martel2012}
Catherine de~Martel, Jacques Ferlay, Silvia Franceschi, J{\'e}r{\^o}me Vignat,
  Freddie Bray, David Forman, and Martyn Plummer.
\newblock {Global burden of cancers attributable to infections in 2008: a
  review and synthetic analysis.}
\newblock {\em Lancet. Oncol.}, 13(6):607--615, June 2012.

\bibitem{Dingli:2007ts}
David Dingli, Arne Traulsen, and Jorge~M Pacheco.
\newblock {Stochastic dynamics of hematopoietic tumor stem cells}.
\newblock {\em Cell Cycle (Georgetown, Tex)}, 6(4):461--466, February 2007.

\bibitem{Dykstra:2007dp}
Brad Dykstra, David Kent, Michelle Bowie, Lindsay McCaffrey, Melisa Hamilton,
  Kristin Lyons, Shang-Jung Lee, Ryan Brinkman, and Connie Eaves.
\newblock {Long-term propagation of distinct hematopoietic differentiation
  programs in vivo.}
\newblock {\em Stem Cell}, 1(2):218--229, August 2007.

\bibitem{Elinav:2013hv}
Eran Elinav, Roni Nowarski, Christoph~A Thaiss, Bo~Hu, Chengcheng Jin, and
  Richard~A Flavell.
\newblock {Inflammation-induced cancer: crosstalk between tumours, immune cells
  and microorganisms.}
\newblock {\em Nat Rev Cancer}, 13(11):759--771, October 2013.

\bibitem{Gibson:2015ky}
Tyler~M Gibson and Charles~A Gersbach.
\newblock {Single-molecule analysis of myocyte differentiation reveals bimodal
  lineage commitment.}
\newblock {\em Integr Biol (Camb)}, 7(6):663--671, June 2015.

\bibitem{Graf:2002tj}
Thomas Graf.
\newblock {Differentiation plasticity of hematopoietic cells.}
\newblock {\em Blood}, 99(9):3089--3101, May 2002.

\bibitem{Grivennikov:2010dk}
Sergei~I Grivennikov, Florian~R Greten, and Michael Karin.
\newblock {Immunity, inflammation, and cancer.}
\newblock {\em Cell}, 140(6):883--899, March 2010.

\bibitem{Guo:2017ix}
Yucheng Guo, Qing Nie, Adam~L MacLean, Yanda Li, Jinzhi Lei, and Shao Li.
\newblock {Multiscale modeling of inflammation-induced tumorigenesis reveals
  competing oncogenic and onco-protective roles for inflammation.}
\newblock {\em Cancer Research}, 77(22):6429--6441, September 2017.

\bibitem{Hanahan:2000hx}
D~Hanahan and R~A Weinberg.
\newblock {The hallmarks of cancer.}
\newblock {\em Cell}, 100(1):57--70, January 2000.

\bibitem{Hanselmann:2016kg}
Rainer~G Hanselmann and Cornelius Welter.
\newblock {Origin of Cancer: An Information, Energy, and Matter Disease.}
\newblock {\em Front Cell Dev Biol}, 4:121, 2016.

\bibitem{Hayashi:2008fu}
Katsuhiko Hayashi, Susana M~Chuva de~Sousa~Lopes, Fuchou Tang, and M~Azim
  Surani.
\newblock {Dynamic equilibrium and heterogeneity of mouse pluripotent stem
  cells with distinct functional and epigenetic states.}
\newblock {\em Stem Cell}, 3(4):391--401, October 2008.

\bibitem{Hu:2012iy}
G~M Hu, C~Y Lee, Y-Y Chen, N~N Pang, and W~J Tzeng.
\newblock {Mathematical model of heterogeneous cancer growth with an autocrine
  signalling pathway.}
\newblock {\em Cell Prolif}, 45(5):445--455, September 2012.

\bibitem{Huang:2017jr}
Rongsheng Huang and Jinzhi Lei.
\newblock {Dynamics of gene expression with positive feedback to histone
  modifications at bivalent domains}.
\newblock {\em Int. J. Mod. Phys. B}, 4:1850075, November 2017.

\bibitem{Huang:2019hn}
Rongsheng Huang, Jinzhi Lei, and {,Zhou Pei-Yuan Center for Applied
  Mathematics, MOE Key Laboratory of Bioinformatics, Tsinghua University,
  Beijing China, 100084}.
\newblock {Cell-type switches induced by stochastic histone modification
  inheritance}.
\newblock {\em Discrete {\&} Continuous Dynamical Systems - B}, 22(11):1--19,
  2019.

\bibitem{Lander:2009fr}
Arthur~D Lander, Kimberly~K. Gokoffski, Frederic Y~M Wan, Qing Nie, and Anne~L.
  Calof.
\newblock {Cell lineages and the logic of proliferative control.}
\newblock {\em PLoS Biology}, 7(1):e15, January 2009.

\bibitem{LeMagnen:2018fx}
Cl{\'e}mentine Le~Magnen, Michael~M Shen, and Cory Abate-Shen.
\newblock {Lineage Plasticity in Cancer Progression and Treatment.}
\newblock {\em Annu Rev Cancer Biol}, 2:271--289, March 2018.

\bibitem{Lei:2019vd}
Jinzhi Lei.
\newblock {A general mathematical framework for understanding the behavior of
  heterogeneous stem cell regeneration}.
\newblock {\em BioRxiv}, 592139:https://doi.org/10.1101/592139, 2019.

\bibitem{2014PNAS..111E.880L}
Jinzhi Lei, Simon~A Levin, and Qing Nie.
\newblock {Mathematical model of adult stem cell regeneration with cross-talk
  between genetic and epigenetic regulation.}
\newblock {\em Proc Natl Acad Sci USA}, 111(10):E880--E887, March 2014.

\bibitem{Jinzhi:2011bl}
Jinzhi Lei and Michael~C Mackey.
\newblock {Multistability in an age-structured model of hematopoiesis: Cyclical
  neutropenia}.
\newblock {\em J Theor Biol}, 270(1):143--153, February 2011.

\bibitem{Mackey:1978vy}
M~C Mackey.
\newblock {Unified hypothesis for the origin of aplastic anemia and periodic
  hematopoiesis}.
\newblock {\em Blood}, 51(5):941--956, May 1978.

\bibitem{Mackey:2001ux}
M~C Mackey.
\newblock {Cell kinetic status of haematopoietic stem cells}.
\newblock {\em Cell Prolif}, 34(2):71--83, April 2001.

\bibitem{Malta:2018ic}
Tathiane~M Malta, Artem Sokolov, Andrew~J Gentles, Tomasz Burzykowski, Laila
  Poisson, John~N Weinstein, Bo{\.{z}}ena Kami{\'{n}}ska, Joerg Huelsken,
  Larsson Omberg, Olivier Gevaert, Antonio Colaprico, Patrycja
  Czerwi{\'{n}}ska, Sylwia Mazurek, Lopa Mishra, Holger Heyn, Alex Krasnitz,
  Andrew~K Godwin, Alexander~J Lazar, {Cancer Genome Atlas Research Network},
  Joshua~M Stuart, Katherine~A Hoadley, Peter~W Laird, Houtan Noushmehr, and
  Maciej Wiznerowicz.
\newblock {Machine Learning Identifies Stemness Features Associated with
  Oncogenic Dedifferentiation.}
\newblock {\em Cell}, 173(2):338--354.e15, April 2018.

\bibitem{2008PLoSO...3.1591M}
Marc Mangel and Michael~B. Bonsall.
\newblock {Phenotypic Evolutionary Models in Stem Cell Biology: Replacement,
  Quiescence, and Variability}.
\newblock {\em PLoS ONE}, 3(2):--, 2008.

\bibitem{Mangel:2012ct}
Marc Mangel and Michael~B. Bonsall.
\newblock {Stem cell biology is population biology: differentiation of
  hematopoietic multipotent progenitors to common lymphoid and myeloid
  progenitors.}
\newblock {\em Theor Biol Med Model}, 10(1):5--5, December 2012.

\bibitem{Parkin2006}
Donald~Maxwell Parkin.
\newblock {The global health burden of infection-associated cancers in the year
  2002}.
\newblock {\em Int. J. Cancer}, 118(12):3030--3044, June 2006.

\bibitem{Probst:2009iq}
Aline~V Probst, Elaine Dunleavy, and Genevi{\`e}ve Almouzni.
\newblock {Epigenetic inheritance during the cell cycle.}
\newblock {\em Nat Rev Mol Cell Biol}, 10(3):192--206, February 2009.

\bibitem{Puram:2017ku}
Sidharth~V Puram, Itay Tirosh, Anuraag~S Parikh, Anoop~P Patel, Keren Yizhak,
  Shawn Gillespie, Christopher Rodman, Christina~L Luo, Edmund~A Mroz, Kevin~S
  Emerick, Daniel~G Deschler, Mark~A Varvares, Ravi Mylvaganam, Orit
  Rozenblatt-Rosen, James~W Rocco, William~C Faquin, Derrick~T Lin, Aviv Regev,
  and Bradley~E Bernstein.
\newblock {Single-Cell Transcriptomic Analysis of Primary and Metastatic Tumor
  Ecosystems in Head and Neck Cancer.}
\newblock {\em Cell}, 171(7):1611--1624.e24, December 2017.

\bibitem{Rockne:2019cd}
Russell~C Rockne, Andrea Hawkins-Daarud, Kristin~R Swanson, James~P Sluka,
  James~A Glazier, Paul Macklin, David~A Hormuth~II, Angela~M Jarrett, Ernesto
  A B~F Lima, J~Tinsley~Oden, George Biros, Thomas~E Yankeelov, Kit Curtius,
  Ibrahim Al~Bakir, Dominik Wodarz, Natalia Komarova, Luis Aparicio, Mykola
  Bordyuh, Raul Rabadan, Stacey~D Finley, Heiko Enderling, Jimmy Caudell,
  Eduardo~G Moros, Alexander R~A Anderson, Robert~A Gatenby, Artem Kaznatcheev,
  Peter Jeavons, Nikhil Krishnan, Julia Pelesko, Raoul~R Wadhwa, Nara Yoon,
  Daniel Nichol, Andriy Marusyk, Michael Hinczewski, and Jacob~G Scott.
\newblock {The 2019 mathematical oncology roadmap}.
\newblock {\em Phys. Biol.}, 16(4):041005--34, July 2019.

\bibitem{RodriguezBrenes:2011hw}
I.A. Rodriguez-Brenes, N.L. Komarova, and D.~Wodarz.
\newblock {Evolutionary dynamics of feedback escape and the development of
  stem-cell{\textendash}driven cancers}.
\newblock {\em Proc Natl Acad Sci USA}, 108(47):18983--18988, 2011.

\bibitem{Schepeler:2014ir}
Troels Schepeler, Mahalia~E Page, and Kim~B Jensen.
\newblock {Heterogeneity and plasticity of epidermal stem cells.}
\newblock {\em Development}, 141(13):2559--2567, June 2014.

\bibitem{SerraCardona:2018bz}
Albert Serra-Cardona and Zhiguo Zhang.
\newblock {Replication-Coupled Nucleosome Assembly in the Passage of Epigenetic
  Information and Cell Identity.}
\newblock {\em Trends Biochem Sci}, 43(2):136--148, February 2018.

\bibitem{Singer:2014eua}
Zakary~S Singer, John Yong, Julia Tischler, Jamie~A Hackett, Alphan Altinok,
  M~Azim Surani, Long Cai, and Michael~B Elowitz.
\newblock {Dynamic heterogeneity and DNA methylation in embryonic stem cells.}
\newblock {\em Mol Cell}, 55(2):319--331, July 2014.

\bibitem{Situ:2017dg}
Qiaojun Situ and Jinzhi Lei.
\newblock {A mathematical model of stem cell regeneration with epigenetic state
  transitions}.
\newblock {\em MBE}, 14(5/6):1379--1397, October 2017.

\bibitem{Song:2018kl}
You Song, Siyu Yang, and Jinzhi Lei.
\newblock {ParaCells: A GPU Architecture for Cell-Centered Models in
  Computational Biology.}
\newblock {\em IEEE/ACM Trans. Comput. Biol. and Bioinf.}, 16(3):994--1006,
  March 2018.

\bibitem{Su:2017kp}
Yapeng Su, Wei Wei, Lidia Robert, Min Xue, Jennifer Tsoi, Angel Garcia-Diaz,
  Blanca Homet~Moreno, Jungwoo Kim, Rachel~H Ng, Jihoon~W Lee, Richard~C Koya,
  Begonya Comin-Anduix, Thomas~G Graeber, Antoni Ribas, and James~R Heath.
\newblock {Single-cell analysis resolves the cell state transition and
  signaling dynamics associated with melanoma drug-induced resistance.}
\newblock {\em Proc Natl Acad Sci USA}, 114(52):13679--13684, December 2017.

\bibitem{Takaoka:2014gg}
Katsuyoshi Takaoka and Hiroshi Hamada.
\newblock {Origin of cellular asymmetries in the pre-implantation mouse embryo:
  a hypothesis.}
\newblock {\em Philos Trans R Soc Lond B Biol Sci}, 369(1657):--, December
  2014.

\bibitem{Tarabichi:2013ev}
M~Tarabichi, A~Antoniou, M~Saiselet, J~M Pita, G~Andry, J~E Dumont, V~Detours,
  and C~Maenhaut.
\newblock {Systems biology of cancer: entropy, disorder, and selection-driven
  evolution to independence, invasion and {\textquotedblleft}swarm
  intelligence{\textquotedblright}}.
\newblock {\em Cancer Metastasis Rev}, 32(3-4):403--421, April 2013.

\bibitem{Traulsen:2013dj}
Arne Traulsen, Tom Lenaerts, Jorge~M Pacheco, and David Dingli.
\newblock {On the dynamics of neutral mutations in a mathematical model for a
  homogeneous stem cell population.}
\newblock {\em Journal of the Royal Society, Interface / the Royal Society},
  10(79):20120810--20120810, January 2013.

\bibitem{Wu:2014gw}
Hao Wu and Yi~Zhang.
\newblock {Reversing DNA methylation: mechanisms, genomics, and biological
  functions.}
\newblock {\em Cell}, 156(1-2):45--68, January 2014.

\bibitem{ZernickaGoetz:2009ita}
Magdalena Zernicka-Goetz, Samantha~A Morris, and Alexander~W Bruce.
\newblock {Making a firm decision: multifaceted regulation of cell fate in the
  early mouse embryo.}
\newblock {\em Nat Rev Genet}, 10(7):467--477, June 2009.

\bibitem{Zhang:2019fs}
Can Zhang, Changyong Shao, Xiaopei Jiao, Yue Bai, Miao Li, Hanping Shi, Jinzhi
  Lei, and Xiaosong Zhong.
\newblock {Leukemic cell plasticity induces immune escape after CD19 chimeric
  antigen receptor T cell therapy of acute B lymphoblastic leukemia}.
\newblock Preprint, 2019.

\bibitem{Zhou:2013ky}
Da~Zhou, Dingming Wu, Zhe Li, Minping Qian, and Michael~Q Zhang.
\newblock {Population dynamics of cancer cells with cell state conversions.}
\newblock {\em Quant Biol}, 1(3):201--208, September 2013.

\end{thebibliography}

\end{document}